\documentclass{ws-p10x7}

\begin{document}
\def\mtiny{\vrule width 0pt}
\def\mrm#1{\mathrm{#1}}
\def\DZ{\relax\ifmmode{D^0}\else{$\mrm{D}^{\mrm{0}}$}\fi}
\def\DZB{\relax\ifmmode{\overline{D}\mtiny^0}
        \else{$\overline{\mrm{D}}\mtiny^{\mrm{0}}$}\fi}
\def\BZ{\relax\ifmmode{B^0_d}\else{$\mrm{B}^{\mrm{0}}_{\mrm{d}}$}\fi}
\def\BZB{\relax\ifmmode{\overline{B}\mtiny^0_d}
        \else{$\overline{\mrm{B}}\mtiny^{\mrm{0}}_{\mrm{d}}$}\fi}
\def\BZS{\relax\ifmmode{B^0_s}\else{$\mrm{D}^{\mrm{0}}_{\mrm{s}}$}\fi}
\def\BZSB{\relax\ifmmode{\overline{B}\mtiny^0_s}
        \else{$\overline{\mrm{D}}\mtiny^{\mrm{0}}_{\mrm{s}}$}\fi}
\def\KZ{\relax\ifmmode{K^0}\else{$\mrm{K}^{\mrm{0}}$}\fi}
\def\KZB{\relax\ifmmode{\overline{K}\mtiny^0}
        \else{$\overline{\mrm{K}}\mtiny^{\mrm{0}}$}\fi}
\title{CLEO Results on Heavy Meson Mixing}

\author{Harry N. Nelson}

\address{Physics Department, University of California, Broida Hall (Bldg. 572), Santa Barbara, CA 93106-9530,
USA\\E-mail: hnn@hep.ucsb.edu}

\twocolumn[\maketitle\abstract{ We discuss recent CLEO results
on $\DZ-\DZB$ and $\BZ-\BZB$ mixing.  The principal results are
that for the $\DZ$ system, allowing for CP violations,
the mixing amplitude $x^\prime<2.9\%$ (95\% C.L.), and for
the $\BZ$ system, $\chi=0.198\!\pm\!0.013\!\pm\!0.014$.  We make
projections for future sensistivity to $\DZ-\DZB$ mixing,
and to $\sin(2\beta+\gamma)$.}]

\section{Introduction}

The $\DZ\!-\!\DZB$ system is unlike other systems that mix,
such as $\KZ\!-\!\KZB$, $\BZ\!-\!\BZB$, and $\BZS\!-\!\BZSB$ in
at least two respects:
first, the Standard Model contributions are thought to be extremely
small, so non-Standard contributions might be obvious; second,
$\DZ\!-\!\DZB$ is the only system that consists of up type quarks.
New physics that
differentiates between up and down type quarks could be revealed
by study of $\DZ\!-\!\DZB$.  Indeed, there are numerous
models, with relevant particles as massive as 100 TeV,
that predict large $\DZ\!-\!\DZB$ mixing.\cite{hnlp}

We describe recent results from CLEO II.V on
$\DZ\!-\!\DZB$ mixing, where we use the sequence 
$D^{*+}\!\to\!\DZ\pi^+_s$, where the charged pion is `slow' in
momentum, followed by the appearence of a $K^+\pi^-$
final state, and the sequence formed by application of charge
conjugations.  We also describe results on $\BZ\!-\!\BZB$
mixing at the $\Upsilon(\mathrm{4s})$.  Following the decay of the
$\Upsilon(\mathrm{4s})$ to $\BZ\BZB$,  we tag one $\BZ$ with a semileptonic
decay, and the other by a partial reconstruction of the exclusive
final state $D^{*+}\pi^-$.

\section{\DZ-\DZB\ Mixing}\label{sec:dd}

The final configuration of the CLEO II detector, known as
CLEO II.V, took 9.0 fb$^{-1}$ of $e^+e^-$ collisions between 1996
and 1999.  

\begin{figure}
\epsfig{figure=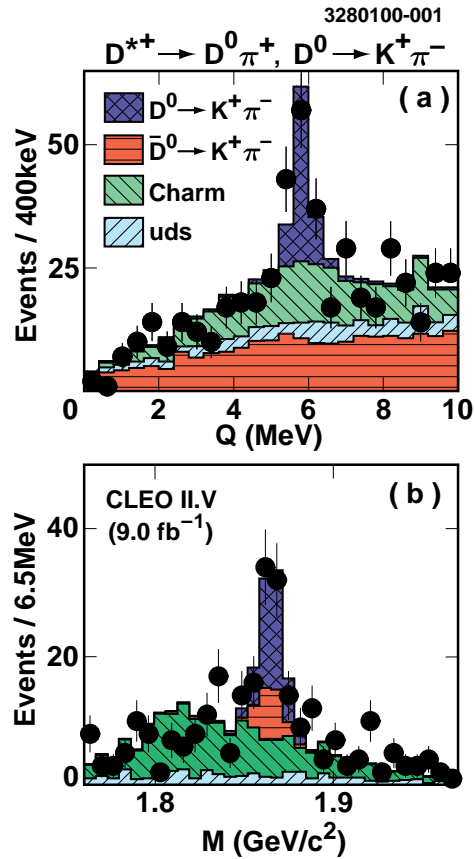,width=15pc}
\caption{Signal for the `wrong sign' decay $D^0\!\to\!K^+\pi^-$,
projected onto $Q$ (a) and $M$ (b).
The fit for the signal and various backgrounds are given by the
hatched and colored histograms.}
\label{fig:wsct}
\end{figure}

CLEO II.V featured a vertex detector (SVX)
consisting of three layers of double-sided silicon, and helium
as the drift chamber gas.  
The measurement of $z$
by the SVX narrowed the resolution in $Q$, the energy released
in the $D^{*+}\!\to\!D^0\pi^+$ decay to $\sigma_Q=190\,$KeV,
or about 1/4 that obtained in earlier CLEO work.  The
use of helium narrowed the resolution in $M$, the mass reconstructed
in $D^0\!\to\!K^{\mp}\pi^{\pm}$, to $\sigma_M=6.4\,$MeV, which
is nearly 1/2 that of earlier CLEO work.  Both of these resolution
improvements are important for detecting the signal of
$D^0\!\to\!K^+\pi^-$, which is shown in Fig.~\ref{fig:wsct}.

The fit in Fig.~\ref{fig:wsct}
indicates $44.8^{+9.7}_{-8.7}$  signal events.\cite{cleows}
The rate of
`wrong-sign' decay, relative to `right-sign decay' is
$0.332^{+0.063}_{-0.065}\pm0.040\,$\%, which is
close to $\tan^4\theta_C$.

\begin{figure}
\epsfig{figure=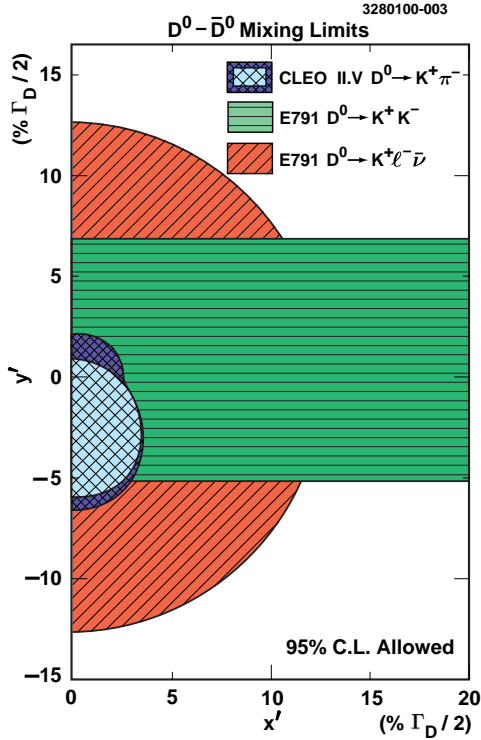,width=15pc}
\caption{Allowed regions for the mixing amplitudes, $x^\prime$ and
$y^\prime$, at the 95\% C.L.  Those nearest the origin are
from this work, where the inner(outer) region requires(does not
require) CP conservation.}
\label{fig:xylim}
\end{figure}

We analyze the decay times of the events in the signal region to
deduce results on the $\DZ\!\to\!\DZB$ mixing amplitudes.  The
normalized amplitude $x(y)$ describes transitions through off(on)-shell
intermediate states.  The existence of a direct decay amplitude
for $\DZ\!\to\!K^+\pi^-$ complicates the analysis.  The direct
decay contributes a purely exponential
 distribution of decay times.
The direct decay amplitude might have a strong phase shift $\delta$
relative to the favored decay amplitude, $\DZB\!\to\!K^+\pi^-$, and
so the interference of mixing and direct decay contributes decay times
according to the distribution $y^\prime t e^{-t}$, where 
$y^\prime=y\cos\delta - x\sin\delta$.  Pure mixing then contributes
decay times according to the distribution $(1/4)(x^2\!+\!y^2)t^2e^{-t}$.

Fits to our data result in the allowed regions in Fig.~\ref{fig:xylim}.
In our principal results we allow  CP violation simultaneously
in all three terms of the time evolution: in direct decay, interference,
and in mixing.

New physics would most probably appear in the amplitude $x$.  When
$\delta=0$, $x^\prime=x$, and our limit is $x<2.9\%$ at 95\% C.L.
At roughly $x\sim1.0\%$, $\DZ\!-\!\DZB$ mixing would surpass 
$\KZ\!-\!\KZB$ mixing as the most tight constraint on flavor changing
neutral currents.

Our technique is now limited by wrong-sign `background' from
the direct decay. We then predict that for future work at
the $B$-factories, this technique will give new sensitivity only
as the one-quarter power of the integrated luminosity, and will
reach about $x^\prime=0.7\%$ at 1000~fb$^{-1}$.  In contrast,
there are techniques that might be background-free at the
$\psi^{\prime\prime}$, and so the scaling with luminosity would
go as the one-half power, and hit about $x=0.1\%$ at 1000~fb$^{-1}$.

\section{\BZ-\BZB\ Mixing}\label{sec:bb}

\begin{figure}
\epsfig{figure=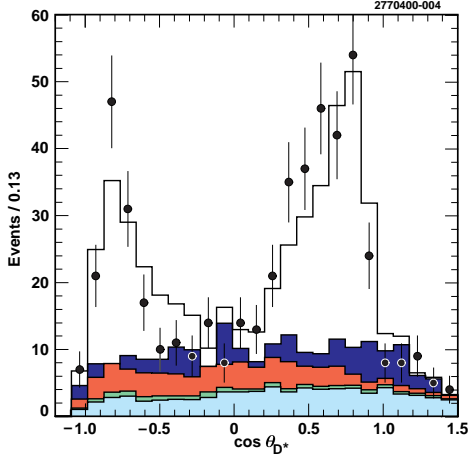,width=15pc}
\caption{Like sign events, which contain the $B$ mixing signal, as a function
of the decay angle of the $D^{*+}$, $\cos{\theta^*}$.
The data are the points, and the fit results are full histograms,
with backgrounds colored.}
\label{fig:bmix}
\end{figure}

At CESR, the $B$'s do not move sufficiently to allow the measurement
of decay times.  Thus, we are sensitive only to the effect of
mixing after integration over the decay time variables, in particular,
$\chi=\Gamma(\BZ\!\to\!\BZB)/[\Gamma(\BZ\!\to\!\BZ)+\Gamma(\BZ\!\to\!\BZB)]
\approx(x^2\!+\!y^2)/2/(1\!+\!x^2)$, where $x$ and $y$ were described
earlier, but in this case are for the $\BZ\!-\!\BZB$ system.

We look for the process $\Upsilon(\mathrm{4s})\!\to\!\BZ\BZB\!\to\!\BZB\BZB$,
followed by one $\BZB\!\to\!X\ell^-\overline{\nu}_{\ell}$, and the other
$\BZB\!\to\!D^{*+}h_W^-$, and the processes obtained by charge
conjugations.  The hadron from the $W^-$, $h_W^-$, can be either a
$\pi^-$ or a $\rho^-$, and is fully reconstructed.  We reconstruct
only the slow pion from $D^{*+}\!\to\!D^0\pi_s^+$; there is sufficient
information to reconstruct all of the decay 
kinematics with zero constraints.\cite{dspi}

The complete CLEO-II data set of 9.1~fb$^{-1}$ taken on the
$\Upsilon(\mathrm{4s})$ resonance is used to measure the
mixing signal, and 4.4~fb$^{-1}$ taken off-resonance is used
to estimate various backgrounds.  We observe 458 mixed or
`like sign' events $(\ell^{\pm}h_W^{\pm})$, shown in Fig.~\ref{fig:bmix},
and 1524 unmixed or `unlike sign' events
$(\ell^{\pm}h_W^{\mp})$.\cite{bmixpre}

After correction, we find $\chi=0.198\pm0.013\pm0.014$.
The principal contributions to the systematic error come
from cases where the $\BZB\!\to\!D^{*+}h_W^-$ decay is
a mis-tagged $\BZ$ decay ($0.009$), charged $B$ background
($0.007$), and uncertainty over two body background ($0.006$).

If we assume that $|y|\ll x$, as is theoretically expected, we
can conclude that $\Delta m=0.523\pm0.029\pm0.031\mathrm{ps}^{-1}$.  
Comparison of the charge states of the like sign events allows 
us to restrict CP violation in $\BZ$ state mixing by
$|\mathrm{Re}(\epsilon_B)|<3.4\%$, 95\% C.L.

At an accelerator where the $B$ decay times can be measured,
the events used here to measure mixing can be used to
measure $\sin(2\beta+\gamma)$.\cite{sintb}  The work in
Ref.~5 omitted a form factor suppression in the
path through Cabibbo-suppressed decay, and so resulted in
an optimistic projection.  Using the results here, and with
improvements expected at the $B$-factories from better tagging
and use of decay time dependence, we can project that an
error on $\sin(2\beta+\gamma)$ of about $1/3$ can be reached
with 200~fb$^{-1}$.

\section*{Acknowledgments}
This work is based on the Ph.~D. dissertations of
David Asner and Andrew Foland.  I would like to thank
the conference organizers for their delightful conference,
and particularly Taku Yamanaka for his extra attention,
and the help desk for guiding me to a bicycle shop.
The author's conference attendance and this work were supported
by the Department of Energy under contract
DE-FG03-91ER40618.

\end{document}